\documentclass[aps,prl,twocolumn,showpacs,groupedaddress]{revtex4-1}
\usepackage{graphicx}  
\usepackage{dcolumn}   
\usepackage{bm}        
\usepackage[english]{babel}
\usepackage{amsfonts,amsmath,amssymb,mathrsfs}
\usepackage{epstopdf}
\usepackage{subfigure}
\usepackage{color}

\def\a{\alpha}
\def\r{\rho}
\def\s{\sigma}
\def\t{\tau}
\def\m{\mu}
\def\n{\nu}
\def\k{\kappa}
\def\th{\theta}
\def\g{\gamma}\def\G{\Gamma}
\def\L{\Lambda}\def\l{\lambda}
\def\D{\Delta}
\def\la{\langle}
\def\ra{\rangle}
\def\o{\omega}\def\O{\Omega}
\def\d{\delta}
\def\p{\partial}

\def\half{\textstyle{\frac{1}{2}}}

\def\bdoc{\begin{document}}
\def\edoc{\end{document}}

\def\beq{\begin{equation}}
\def\eeq{\end{equation}}
\def\bea{\begin{eqnarray}}
\def\eea{\end{eqnarray}}
\def\ben{\begin{enumerate}}
\def\een{\end{enumerate}}
\def\la{\langle}
\def\ra{\rangle}
\def\a{\alpha}
\def\b{\beta}
\def\g{\gamma}\def\G{\Gamma}
\def\d{\delta}\def\D{\Delta}
\def\e{\epsilon}

\def\th{\theta}
\def\k{\kappa}
\def\l{\lambda}
\def\m{\mu}
\def\n{\nu}
\def\o{\omega}
\def\p{\pi}
\def\r{\rho}
\def\s{\sigma}
\def\t{\tau}
\def\L{{\cal L}}
\def\S{\Sigma }
\def\gsim{\; \raisebox{-.8ex}{$\stackrel{\textstyle >}{\sim}$}\;}
\def\lsim{\; \raisebox{-.8ex}{$\stackrel{\textstyle <}{\sim}$}\;}
\def\gtrsim{\gsim}
\def\lessim{\lsim}
\def\loc{{\rm local}}
\def\vm{v_{\rm max}}
\def\bh{\bar{h}}
\def\del{\partial}
\def\nab{\nabla}
\def\half{{\textstyle{\frac{1}{2}}}}
\def\fourth{{\textstyle{\frac{1}{4}}}}

\def\bD{{\bf D}}
\def\bE{{\bf E}}
\def\bF{{\bf F}}
\def\bB{{\bf B}}
\def\bP{{\bf P}}
\def\bV{{\bf v}}
\def\bv{{\bf v}}
\def\bx{{\bf x}}
\def\by{{\bf y}}
\def\bz{{\bf z}}
\def\ba{{\bf a}}
\def\bd{{\bf d}}
\def\bs{{\bf s}}
\def\bn{{\bf n}}
\def\bp{{\bf p}}

\def\O{\Omega}

\def\br{{\bf r}}
\def\bnab{{\bf \nab}}

\def\tE{\tilde{E}}
\def\tL{\tilde{L}}

\begin{document}

\title{Spinning Black Holes as Particle Accelerators}
\author{Ted Jacobson$^*$ and Thomas P. Sotiriou$^{\dagger}$}
\affiliation{$^*$Center for Fundamental Physics,  University of Maryland, College Park, MD 20742-4111, USA}
\affiliation{$^\dagger$Department of Applied Mathematics and Theoretical Physics, Centre for  
Mathematical Sciences, University of Cambridge, Wilberforce Road,  
Cambridge, CB3 0WA, UK}
\date{\today} 
\begin{abstract}
It has recently been pointed out that  particles falling freely from  
rest at infinity outside a Kerr black hole can in principle collide with  
arbitrarily high center of mass energy
in the limiting case of maximal black hole spin.
Here we aim to elucidate the mechanism for this fascinating result,  
and to point out its practical limitations, which imply that ultra-energetic  
collisions cannot occur near black holes in nature.

\end{abstract}  
\pacs{04.70.Bw, 04.70.-s, 97.60.Lf}
\maketitle


Ba\~{n}ados, Silk and West (BSW)~\cite{Banados:2009pr}
recently showed that
particles falling freely from rest outside a Kerr black hole can collide with arbitrarily high
center of mass energy in the limiting case of maximal black hole spin.
They proposed that this might lead to signals from ultra high energy collisions,
for example of  dark matter particles. 
In this letter we aim to elucidate the mechanism for this result, and to point
out its practical limitations given that extremal black holes do not exist in nature.
In particular, we clarify why infinite collision energy can only be attained at the horizon, 
and with a maximally spinning black hole. We also show that the maximum 
center of mass energy grows very slowly as the black hole spin approaches its 
maximal value, so it will not be so high for astrophysical black holes.
Finally we calculate the upper bound for the energy of the ejecta of the collision and 
find that to be only slightly above the mass of the particles, even in the extremal limit.
We use units with $G=c=M=1$, 
where $M$ is the black hole mass, 
and metric signature $(+{-}{-}{-})$.

While one can theoretically  extract 100\% of the rest energy of a  
mass
by lowering it into a nonrotating black hole, and one can extract
even more energy using a Penrose process lowering it into
a rotating black hole, neither of these possibilities suggests
that just by falling in freely from far away, a pair of particles
can experience an infinite collision energy
in their center of mass frame. If this
is indeed possible then
although
the debris would be redshifted on the way out, 
it might still reveal features of the S-matrix at arbitrarily high
energies.
This is surprising since one seems to get an infinite 
energy boost---despite conservation of energy---from the finite 
process of falling into the black hole.
But this is a misconception, as we will explain, since it takes in  
fact an infinite time
to access the infinite collision energy.

We restrict attention here to orbits in the equatorial plane of a 
Kerr black hole 
with spin parameter $a$. Given the
energy $E$, angular momentum $l$, and the unit 4-velocity  
condition, one can solve for the four velocity $u$ at any given 
(Boyer-Lindquist) 
radial coordinate $r$, up to a discrete
ambiguity in the sign of $\dot{r}$. Then
one can compute the squared center of mass energy for a pair of particles of mass $m$, 
\begin{equation}
E_{\rm cm}^2= (mu_1 + mu_2)^2 = 2m^2(1 + u_1\cdot u_2),
\end{equation}
where the square and dot refer to the local Lorentz metric. For the case that the particles 
begin at rest at infinity, $E=m$, this yields
equation (8) of  Ref.~\cite{Banados:2009pr},
%
\bea
&&\Big(E^{^{Kerr}}_{\rm cm}\Big)^2=\frac{2\, m^2}{r (r^2 - 2r + a^2)}\times\nonumber\\
&&\!\!\!\!\!\Big(  2 a^2 (1 + r)-2 a (l_2 + l_1) - l_2 l_1 (-2 + r) + 2 (-1 + r) r^2  \nonumber\\ 
 && \!\!\!\!\!-\sqrt{2 (a - l_2)^2 - l_2^2 r + 2 r^2} \sqrt{2 (a - l_1)^2 - l_1^2 r + 2 r^2}\, \Big).\label{EcmK}
\eea

The largest collision energy for such particles 
occurs when they collide at the horizon, carrying the maximum 
and minimum angular momenta that permit a fall all the way to the horizon.
(We have not proved this analytically, but rather by numerical exploration.)
These angular momenta correspond to those at which the centrifugal 
barrier drops just low enough so that there is no turning point for the radial
motion. The particles therefore fall on a trajectory that spirals asymptotically 
into an unstable circular orbit at some critical radius, taking a logarithmically 
divergent proper time to do so. Another branch of the trajectories begins at this
orbit and spirals into the black hole. It is this latter branch on which the maximum
collision energy occurs at the horizon. 

The location of the critical radius can be found 
using the effective potential for the radial motion
with unit Killing energy per unit mass
in the equatorial plane.
The proper time derivative
of the (Boyer-Lindquist) radial coordinate of orbital motion 
satisfies 
\beq
\dot{r}^2/2 + V_{\rm eff}(r,l)=0,
\eeq
where the effective potential is given in terms of the 
angular momentum $l$ per unit mass by~\cite{Wald:1984rg}
\beq
V_{\rm eff}= -\frac{1}{r}+\frac{l^2}{2r^2}
-\frac{(l-a)^2}{r^3}.
\eeq
The critical point we are looking for is defined by 
\beq
V_{\rm eff}= dV_{\rm eff}/dr= 0,
\eeq
and is found to occur with
angular momenta 
\beq
l=l_\pm=\pm2(1 +\sqrt{1\mp a}),
\eeq
and at radius
\beq
r=r_\pm=2\mp a+2\sqrt{1\mp a}.
\eeq
In the nonspinning case this yields $l_\pm=\pm4$ and $r_\pm=4$, which lies
well-separated from the horizon. 
With these values, (\ref{EcmK})
gives $E^{Kerr}_{\rm cm}=2\sqrt{2}m$, 
while at the horizon these same angular momenta give 
$E^{Kerr}_{\rm cm}=2\sqrt{5}m$~\cite{Baushev:2008yz}.
In the maximally spinning case it yields $l_\pm=2, 2(1+\sqrt{2})$ 
and $r_\pm=1, (3+2\sqrt{2})$. The horizon lies at 
$r_h=(1+\sqrt{1-a^2})$,  so that in the extremal case $a=1$, 
the critical radius $r_+=1$ coincides with the horizon. 

In the maximally spinning case the 
corotating critical orbit is thus asymptotically
tangent to the horizon. 
Its 4-velocity therefore tends to the null direction generating the horizon, 
since any other direction in the horizon is spacelike. In other words,
the particle is moving at the speed of light, so the center of mass energy 
with any particle not on this horizon generator is infinite. 
This makes clear why infinite collision energy can only 
be attained at the horizon, and with a maximally spinning black hole. 
Since the particle never crosses the horizon, an infinite 
proper time passes for the particle as it spirals asymptotically onto the horizon.
The nature of the divergence can be seen from the radial equation
$\dot{r} = \sqrt{-2V_{\rm eff}(r)}$. At the critical orbit radius $r_\pm$ 
the effective potential has a maximum and vanishes; hence nearby 
it is a negative quadratic function, 
$V_{\rm eff}=-r_\pm^{-3}(r-r_\pm)^2 +\cdots$. 
Therefore
$\dot{r}\propto (r-r_\pm)$, so the proper time diverges
logarithmically as $r_\pm$ is approached. 

As BSW pointed out,  
for a black hole with spin parameter $a$ less than $M$ there 
will be an upper bound to the energy. What is somewhat surprising 
is how slowly the largest collision energy grows as the maximally
spinning case is approached. 
We can estimate the maximal energy with the help of
eq.~(\ref{EcmK}). 
In terms of the small parameter $\e=1-a$, we find that the 
maximal collision energy per unit mass, i.e. the relative 
gamma factor, 
is approximately 
\beq
\frac{E^{max}_{\rm cm}}{m}\sim4.06\e^{-1/4} + O(\e^{1/4}).
\eeq
 In particular, for 
$\e=0.1,\, 0.01,\, 0.001,\, 0.0001$ the numerical result
is, respectively, 
$6.90, 12.5, 22.6, 40.5$. 
For an astrophysical 
black hole, accretion processes prohibit  
any spin factor greater than $a=0.998$ as a theoretical upper limit
\cite{Thorne:1974ve}, and MHD 
simulations~\cite{Gammie:2003qi} 
suggest the
smaller upper
limit of $a\lessim 0.95$. These imply upper bounds 
of around 20 and 10 respectively for the maximum collision 
gamma factor. Hence it seems
that, even neglecting the effects of gravitational radiation~\cite{Berti:2009bk}, 
hyper-relativistic
collision energies will not be realized in nature. 

Note that the essential ingredient in the large collision energy is that one of the particles
has the maximum angular momentum $l_+$. Above we indicated the result if the
other particle has the minimum angular momentum $l_-$ 
and the collision occurs at
the horizon. If instead the collision occurs at $r_+$ (the critical unstable corotating circular orbit)
the coefficient $4.06$ of $\e^{-1/4}$ in the leading approximation is replaced by
$3.70$. If the collision occurs at the horizon, but the second particle
has zero angular momentum, it is replaced by $2.20$. And if the
collision occurs at $r_+$, and the second particle has zero angular momentum,
it is replaced by 2.00. These examples illustrate that, if
the collision energy is to be much larger for a spinning black
hole than in the nonspinning case, the key is for one of the particles
to carry the maximum angular momentum that can be captured.
The angular momentum of the other particle is not really constrained,
nor is the location of the collision, as long as it lies at or
inside the critical radius $r_+$.

Finally, another point made by BSW is that although the 
collision energy can be arbitrarily large in the 
extremal limit, the energy of any collision products ejected to
infinity will be redshifted. We can obtain an upper bound for the
ejecta energy from a collision at the horizon
as follows. In this limiting case 
one of the particles has a 4-momentum vector $k$ that is 
(asymptotically) tangent to
the horizon generator, while the other particle, 
has 4-momentum $p$. 
In order not to fall into the black hole,
the 4-momentum of an ejecta particle must also be tangent to the
horizon generator, so is $\l k$ for some scalar $\l$.
(This is the marginal case. 
To escape, a particle must start strictly outside
the horizon.) If the remaining
reaction products have total 4-momentum $p'$, then
\beq
p+k=p'+\l k,
\eeq
hence 
\beq
p'=p+ (1-\l)k.
\eeq
 Now since $p$, $p'$, and $k$  are all future pointing
vectors, $p'\cdot p>0$ and $k\cdot p>0$, hence $\l-1 <p\cdot p/k\cdot p$.
If the mass of each of the initial particles is $m$, we have 
\beq
 \frac{p\cdot p}{k\cdot p}=\frac{m^2}{(E_{\rm cm}^2/2 - m^2)} 
 \eeq
 Thus
 the ejecta particle can have Killing energy 
 at most twice
 that of $k$, i.e.\ at most $2m$. As the collision energy increases, this ejecta
 energy drops to something just slightly above $m$.

To summarize, we have examined 
some 
practical limitations of using black holes as particle accelerators, 
as proposed in Ref.~\cite{Banados:2009pr}. Infinite center of mass 
energies for the colliding particles can only be attained when the 
black hole is exactly extremal and only at infinite time and on the 
horizon of the black hole. Additionally, the upper bound on the 
collision energy for an astrophysically realistic black hole 
is, neglecting radiation,
less than ten times
the mass of the particles. The
energy of the ejecta of the collision
is no more than twice the particle rest mass.
In conclusion, 
intriguing as it may be in principle, 
the possibility of spinning black holes catalyzing  
hyper-relativistic particle collisions does not seem realizable in practice.

We thank M. Ba\~nados, J. Silk and S. West for helpful discussions. 
This research was supported in part by the NSF
under Grant No.~PHY-0903572, and by STFC. \\

{\em Note added:} As the present manuscript was being completed,
a comment by Berti et al~\cite{Berti:2009bk} appeared which also presents
some of the points made here.

{\em Note added in proof:} The possibility of ultrahigh energy
collisions catalyzed by a rotating black hole
was noticed long ago, in the context of
the study of collisional Penrose processes \cite{PSK, Piran:1977dm}.
The energy that can be extracted at
infinity was analyzed there as well.

 \edoc
\begin{thebibliography}{100}

  
\bibitem{Banados:2009pr}
  M.~Banados, J.~Silk and S.~M.~West,
  ``Kerr Black Holes as Particle Accelerators to Arbitrarily High  
Energy,''
  Phys.\ Rev.\ Lett.\  {\bf 103}, 111102 (2009)
  [arXiv:0909.0169 [hep-ph]].
  
     
\bibitem{Wald:1984rg}
  R.~M.~Wald,
  {\it General Relativity}
(Univ. Chicago Press, Chicago, 1984).
  
\bibitem{Baushev:2008yz}
  A.~N.~Baushev, Int.\ J.\ Mod.\ Phys.\ D {\bf 18}, 1195 (2009)
  ``Dark matter annihilation in the gravitational field of a black hole,''
 [ arXiv:0805.0124 [astro-ph]].

\bibitem{Thorne:1974ve}
  K.~S.~Thorne,
  ``Disk Accretion Onto A Black Hole. 2. Evolution Of The Hole,''
  Astrophys.\ J.\  {\bf 191}, 507 (1974).
  
\bibitem{Gammie:2003qi}
  C.~F.~Gammie, S.~L.~Shapiro and J.~C.~McKinney,
  ``Black Hole Spin Evolution,''
  Astrophys.\ J.\  {\bf 602}, 312 (2004)
  [arXiv:astro-ph/0310886].
  
\bibitem{Berti:2009bk}
  E.~Berti, V.~Cardoso, L.~Gualtieri, F.~Pretorius and U.~Sperhake,
  ``Comment on 'Kerr Black Holes as Particle Accelerators to Arbitrarily High
  Energy','' Phys.\ Rev.\ Lett.\ {\bf 103}, 239001 (2009)
  [arXiv:0911.2243 [gr-qc]].
  
\bibitem{PSK}
T.~Piran, J.~Shaham, and J.~Katz,
``High Efficiency of the Penrose Mechanism for Particle Collisions,"
Astrophys.\ J.\ Lett.\ {\bf 196}, L107 (1975).

\bibitem{Piran:1977dm}
 T.~Piran and J.~Shaham,
 ``Upper Bounds On Collisional Penrose Processes Near Rotating Black Hole
 Horizons,''
 Phys.\ Rev.\  D {\bf 16}, 1615 (1977).


  
 %
\end{thebibliography}
